\documentclass[doublecol,figures]{epl2}
\usepackage{amsmath,amssymb}
\usepackage{color}

\title{Network evolution towards optimal dynamical performance}

\author{Steffen Karalus \and Markus Porto}
\institute{Institut f\"ur Theoretische Physik, Universit\"at zu K\"oln - Z\"ulpicher Str.~77, 50937~K\"oln, Germany, EU}


\abstract{
  Understanding the mutual interdependence between the behavior of dynamical 
  processes on networks and the underlying topologies promises new insight 
  for a large class of empirical networks.  We present a generic approach to 
  investigate this relationship which is applicable to a wide class of dynamics, 
  namely to evolve networks using a performance measure based on the whole 
  spectrum of the dynamics' time evolution operator.  As an example, we consider the 
  graph Laplacian describing diffusion processes, and we evolve the network structure 
  such that a given sub-diffusive behavior emerges.
}

\begin{document}

\maketitle


\section{Introduction}  The importance of networks as a fruitful modeling
approach to very diverse scientific areas is nowadays well established.  
While initially the network topologies were in the focus of investigation, 
presently the main interest lies in dynamics taking place on networks---in 
the sense that the network defines the interaction pattern between the 
individual elements of the dynamical process~\cite{strogatz_exploring_2001,
albert_statistical_2002,dorogovtsev_evolution_2002,newman_structure_2003,
barrat_dynamical_2008,newman_networks_2010}.  Additionally, networks are 
commonly evolving over time and alter their structure due to random mutations 
or directed changes, growth, or shrinkage, modifying the dynamical behavior.  
Since in most cases the functionality of a network is provided by the dynamics, 
evolution favors structures with a better dynamical performance and eventually 
adapts towards an optimal topology.  A deeper understanding of this mutual 
relation between structure and dynamics can be achieved either by studying 
empirical systems directly, in a biological context for instance physiological 
networks~\cite{bashan_network_2012}, food webs~\cite{martinez_network_2006}, 
or protein interaction networks~\cite{maslov_specificity_2002}, or by modeling 
of network evolution towards a desired dynamical performance and in this way 
gain new insight into many empirical networks.  Concerning the latter, 
the evolution of dynamical networks was previously applied in the contexts of 
modularity in changing environments~\cite{kashtan_spontaneous_2005}, Boolean 
(threshold) dynamics~\cite{bornholdt_topological_2000,oikonomou_effects_2006,
braunewell_reliability_2008,szejka_evolution_2010,greenbury_effect_2010,
shao_dynamics-driven_2009}, and synchronization of oscillatory 
systems~\cite{donetti_entangled_2005,donetti_optimal_2006,donetti_network_2008,
rad_efficient_2008,gorochowski_evolving_2010}.  It is known that the dynamical 
behavior is closely related to the network's spectral 
properties~\cite{atay_network_2006}, which was exploited in studies in which 
the Laplacian eigenratio~\cite{arenas_synchronization_2008} was maximized by 
network evolution~\cite{donetti_entangled_2005,donetti_optimal_2006,
donetti_network_2008,rad_efficient_2008}.  Other studies carried out the 
dynamics explicitly~\cite{bornholdt_topological_2000,oikonomou_effects_2006,
braunewell_reliability_2008,szejka_evolution_2010,greenbury_effect_2010,
gorochowski_evolving_2010}, which can be computationally very costly.

In this Letter, we present an alternative approach based on the whole eigenvalue 
spectrum of the time evolution operator~$\mathcal{O}_{\mathbf{A}}$ of linear dynamics.  
Specifically, we consider dynamics on a network with $N$ vertices described by 
a linear equation,
\begin{equation}
  \label{eq:dynamic}
  \frac{\mathrm{d}\mathbf{p}(t)}{\mathrm{d}t} =
  \mathcal{O}_{\mathbf{A}} \, \mathbf{p}(t) \,,
\end{equation}
where $\mathbf{p}(t) = (p_1(t), \ldots , p_N(t))^\mathrm{T}$ represents the dynamical 
state at time $t$.  Depending on the problem under consideration, $\mathbf{p}(t)$ 
can denote the activity of the individual vertices, their phases, the distribution 
of a substance flowing across the network, the probability distribution of some 
event to occur, etc.  The operator $\mathcal{O}_{\mathbf{A}}$ depends on the network 
topology described by the $N \times N$ adjacency matrix~$\mathbf{A}$ of the network 
(with elements $A_{ij} = 1$ if vertex $i$ is linked to vertex $j$ and $A_{ij} = 0$ 
otherwise).  In contrast to studies on coevolutionary 
networks~\cite{gross_adaptive_2008}, we regard the time scales of the dynamics on 
the network and of the structure modifications as being well separated, which is 
the case in many empirical networks (consider for example cellular processes such 
as protein-protein interaction or gene regulation compared to biological evolution).

\section{Model}  The eigenvalues of $\mathcal{O}_{\mathbf{A}}$, $\lambda_\nu$, 
can be represented in a functional form by the integrated density of states (DOS),
\begin{equation}
  I(\lambda) = \frac{1}{N} \, \sum_{\nu=1}^N \Theta(\lambda - \lambda_{\nu}) \,,
\end{equation}
where $\Theta$ denotes the Heaviside step function, $\Theta(x) = 1$ for $x \ge 0$ and 
$\Theta(x) = 0$ for $x < 0$.  Our approach to quantify the dynamical performance via 
the eigenvalue spectrum is to establish a distance metric $\Delta(I,I^{\mathrm{target}})$ 
between any $I$ and a predefined (but not necessarily realizable) $I^{\mathrm{target}}$ 
yielding the desired dynamical behavior.  The evolution of a network structure consists 
of two parts, carried out in each evolution step $n$.  First, a candidate network is 
constructed from the present one by a random modification in the topology (mutation).  
Secondly, the candidate is accepted or rejected for the next step by a criterion based 
on the values of $\Delta(I,I^{\mathrm{target}})$ before and after the mutation (selection).  
If the distance metric and the criteria for mutation and selection are chosen 
appropriately, the evolution will eventually converge to an optimal network structure 
for a given $\mathcal{O}_{\mathbf{A}}$ and desired dynamical behavior.

In the following discussion, we exemplify our approach by considering simple graphs 
and the graph Laplacian $\mathbf{L}$, given by
\begin{equation}
  L_{ij} = \delta_{ij} k_i - A_{ij} = 
  \begin{cases}
    k_i & \textrm{if } i = j \,, \\
    -A_{ij} & \textrm{if } i \neq j \,,
  \end{cases}
\end{equation}
with $\delta_{ij}$ being Kronecker's delta and $k_i \equiv \sum_j A_{ij}$ the degree 
of vertex~$i$, as dynamical operator, i.e., $\mathcal{O}_{\mathbf{A}} = -C \mathbf{L}$ 
with $C > 0$ being a (diffusion) constant.  The graph Laplacian~$\mathbf{L}$ is of 
importance in many physical situations~\cite{samukhin_laplacian_2008}, such as 
electrical or other flow networks~\cite{gallos_scaling_2007}, Gaussian spring networks 
as polymer models~\cite{gurtovenko_generalized_2005}, random 
walks~\cite{noh_random_2004} as the space discrete equivalent of diffusion, and 
synchronization~\cite{motter_bounding_2007}.  In the context of diffusion, 
$\mathbf{p}(t)$ denotes the probability to find the random walker at a given vertex at 
time~$t$ and eq.~\eqref{eq:dynamic} is the master equation describing its time 
evolution.  Many properties of the random walk, such as the average probability of 
returning to the origin at 
time~$t$, $P_0(t) = N^{-1} \sum_{i=1}^N \exp(-\lambda_{\nu} \, t)$, are determined by the 
spectral dimension~$d_{\mathrm{s}}$ defined via the scaling of the integrated 
DOS~\cite{havlin_diffusion_2002}, 
\begin{equation}
  I(\lambda) \propto \lambda^{d_\mathrm{s} / 2} \,.
  \label{eq:dos}
\end{equation}
It is known that $P_0(t)$ decays for $t \to \infty$ as a power-law, 
$P_0(t) \propto t^{-d_{\mathrm{s}}/2}$~\cite{redner_guide_2001}.  For all uniform Euclidean 
systems without broad waiting time distributions, the mean-square displacement of the 
random walk scales linear with $t$ and $P_0(t)$ decays as $t^{-d/2}$ where $d$ is the 
spatial dimension.  This behavior is called normal diffusion and the corresponding 
spectral dimension is $d_{\mathrm{s}}^{\mathrm{(n)}} = d$, whereas non-linear scaling is 
referred to as anomalous diffusion~\cite{havlin_diffusion_2002}.  An interesting and 
challenging goal is to optimize a network structure for a given fixed 
$1 < d_{\mathrm{s}} < 2$, such that $P_0(t)$ decays slower than in the regular 
two-dimensional (2d) case (sub-diffusion), which will be the desired dynamical behavior 
in the following.

In the present case of the graph Laplacian, there are two issues to be addressed.  
First, the scale of the Laplacian eigenvalues is to first order given by the 
degrees~$\{ k_i \}$ as largest elements of $\mathbf{L}$, located along the diagonal, 
whereas all non-zero off-diagonal elements are $-1$.  Hence, we consider the rescaled 
eigenvalues $\tilde{\lambda}_{\nu} = \lambda_{\nu}/\max_{\nu'} \{ \lambda_{\nu'} \}$, 
which reduces the dependence of the eigenvalue spectrum on the degrees and does not 
change the scaling of $I(\lambda)$.  Second, since the spectral dimension~$d_{\mathrm{s}}$ 
appears in the exponent of eq.~(\ref{eq:dos}), we introduce the logarithmically 
integrated DOS,
\begin{equation}
  \tilde{I}(\ln\tilde{\lambda}) =
  \ln\left(
    \frac{1}{N} \,
    \sum_{\nu=1}^N \Theta(\ln\tilde{\lambda}-\ln\tilde{\lambda}_{\nu})
  \right) \,.
\end{equation}
The measure of dynamical performance can then be defined by the metric
\begin{equation}
  \Delta(\tilde{I},\tilde{I}^{\textrm{target}}) =
  \int_{\ln\tilde{\lambda}_{\mathrm{min}}}^0 \, \left|
    \tilde{I}(\ln\tilde{\lambda}) -
    \tilde{I}^{\textrm{target}}(\ln\tilde{\lambda})
  \right|^2 \, \mathrm{d}\ln\tilde{\lambda}
\end{equation}
where $\tilde{I}$ is the present and $\tilde{I}^{\textrm{target}}$ is the target
logarithmically integrated DOS.  The lower integration boundary
$\ln\tilde{\lambda}_{\mathrm{min}}$ is chosen such that
$\tilde{I}^{\textrm{target}}(\ln\tilde{\lambda}) < \ln(N^{-1})$ for
$\ln\tilde{\lambda} < \ln\tilde{\lambda}_{\mathrm{min}}$ and
$\tilde{I}^{\textrm{target}}(\ln\tilde{\lambda}) \ge \ln(N^{-1})$ for
$\ln\tilde{\lambda} \ge \ln\tilde{\lambda}_{\mathrm{min}}$.  (Note that 
$\tilde{I}^{\textrm{target}}(\ln\tilde{\lambda})$ is a monotonic, but not 
necessarily continuous function; if 
$\tilde{I}^{\textrm{target}}(\ln\tilde{\lambda})$ is continuous, then 
$\tilde{I}^{\textrm{target}}(\ln\tilde{\lambda}_{\mathrm{min}}) = \ln(N^{-1})$.)

Concerning the basic evolution step, we have chosen the mutation such that a 
minimal change in the eigenvalue spectrum is ensured.  A randomly chosen edge 
is removed from the network and a new edge between two previously unconnected vertices 
is introduced.  The selection consist of two parts.  If the modification 
separates the network into two disconnected components then it is rejected directly.  
Otherwise, the mutation is accepted if it decreases the value of the metric 
$\Delta(\tilde{I},\tilde{I}^{\textrm{target}})$.  In this way, the evolving network performs 
an adaptive walk in the space of connected simple graphs with both numbers of 
vertices~$N$ and edges~$M$ fixed.

\section{Results}  As a first test of our evolutionary algorithm, we specify the 
target integrated DOS as~$I^{\textrm{target}}(\lambda) \propto \lambda^{d_{\mathrm{s}}/2}$ 
yielding 
$\tilde{I}^{\textrm{target}}(\ln\tilde{\lambda}) = d_{\mathrm{s}}/2 \, \ln\tilde{\lambda}$, 
with two exemplary values $d_{\mathrm{s}}^{(1)} = 1.4$ and $d_{\mathrm{s}}^{(2)} = 1.1$ both 
in the sub-diffusive regime well below $d_{\mathrm{s}}^{\mathrm{(n)}} = 2$ for a two-dimensional 
system.  In order to verify to which extent the outcome of the evolution depends on the 
initial conditions, i.e., how much of the evolutionary history remains imprinted in the 
networks, we start the evolution from two classes of initial networks, (i)~2d square 
lattices with periodic boundary conditions; and (ii)~connected random graphs $G(N,M)$ 
(simple graphs with $N$ vertices and $M$ randomly chosen edges, restricted to form a 
single connected component).  While in 2d square lattices all vertices have the same 
degree $k=4$, random graphs $G(N,M)$, just as the famous Erd\H{o}s-R\'enyi networks, obey 
a Poisson degree distribution, $P(k) = \eta^k \mathrm{e}^{-\eta}/k!$, with mean degree 
$\langle k \rangle \equiv \eta = 4$ here.  All inital networks are chosen to have 
$N = 361$ vertices.  Each evolution is driven for $10^6$ steps and repeated with different 
random seeds for $100$ realizations.

\begin{figure}
  \centering
  \includegraphics[width=\linewidth]{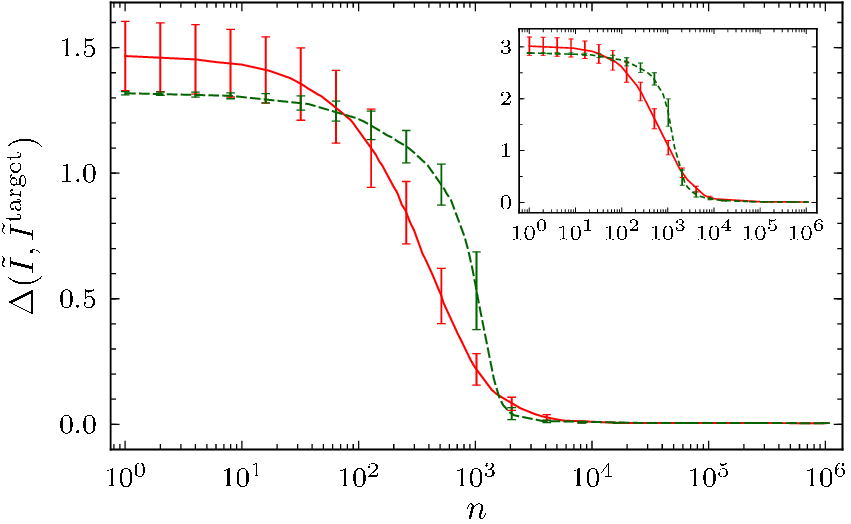}
  \caption{(Color online) Distribution of the distance
    measures~$\Delta(\tilde{I},\tilde{I}^{\textrm{target}})$ as a function of the 
    step number~$n$ evolving with a power-law DOS
    $\tilde{I}^{\textrm{target}}(\ln\tilde{\lambda}) = d_{\mathrm{s}}/2 \,
    \ln\tilde{\lambda}$ with $d_{\mathrm{s}}^{(1)} = 1.4$ (main plot) and
    $d_{\mathrm{s}}^{(2)} = 1.1$ (inset) as target.  The lines display 
    the mean distance measure of $100$ realizations 
    of the evolution starting from 2d square lattices 
    (green dashed lines) and 
    random graphs 
    (red solid lines), and the error bars mark one standard 
    deviation of the respective distributions.}
  \label{fig:score_time}
\end{figure}
In fig.~\ref{fig:score_time}, the distributions of the distance
measure~$\Delta(\tilde{I},\tilde{I}^{\textrm{target}})$ are shown as a function of
the step number~$n$ for the evolution starting with networks from the two
classes of initial networks and evolving towards $d_{\mathrm{s}}^{(1)} = 1.4$ (main plot) 
and $d_{\mathrm{s}}^{(2)} = 1.1$ (inset).  In all cases, the distance measure 
$\Delta(\tilde{I},\tilde{I}^{\textrm{target}})$ decreases slowly for about $10^2$ steps, 
then drops quickly by an order of magnitude within the following decade, and 
afterwards continues to decrease only marginally.  The distributions of the distance 
measure~$\Delta(\tilde{I},\tilde{I}^{\textrm{target}})$ for the two choices of initial 
conditions become indistinguishable after around $10^4$ evolution steps.

\begin{figure}
  \centering
  \includegraphics[width=\linewidth]{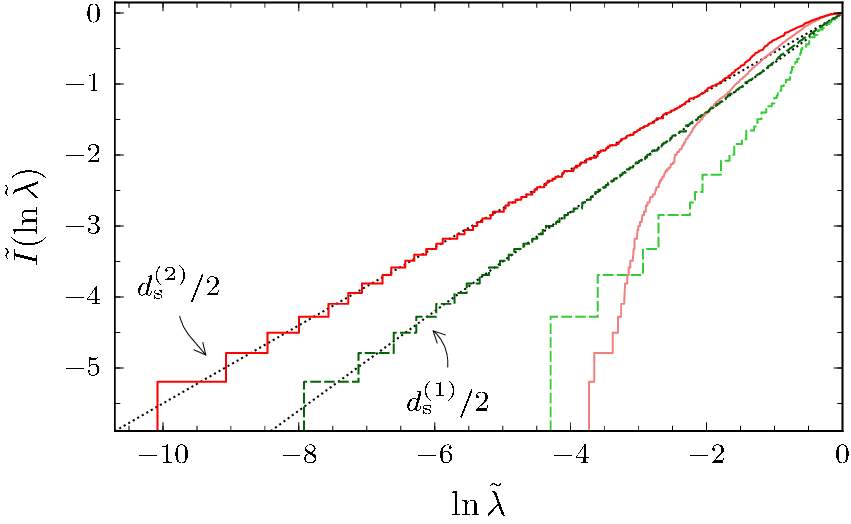}
  \caption{(Color online) Logarithmically integrated 
    DOS~$\tilde{I}(\ln\tilde{\lambda})$ exemplified using individual realizations, 
    two initial networks (light colors) and two networks after $10^6$ evolution 
    steps with different $\tilde{I}^{\textrm{target}}(\ln\tilde{\lambda})$ 
    (dark colors).  Results are shown for evolutions with $d_{\mathrm{s}}^{(1)} = 1.4$ 
    starting from a 2d square lattice 
    (green dashed lines) and $d_{\mathrm{s}}^{(2)} = 1.1$ starting from a 
    random graph (red solid lines).  The two target logarithmically integrated 
    DOS~$\tilde{I}^{\textrm{target}}(\ln\tilde{\lambda}) = 
    d_{\mathrm{s}}/2 \, \ln\tilde{\lambda}$ are shown as references 
    (black dotted lines) with annotated slopes.
  }
  \label{fig:score_snapshot}
\end{figure}
Typical realizations of the logarithmically integrated DOS, as occurring during the 
evolution process shown in fig.~\ref{fig:score_time}, are depicted in 
fig.~\ref{fig:score_snapshot}.  We display four examples, two initial networks and two 
networks at step~$n = 10^6$, having been evolved with target $d_{\mathrm{s}}^{(1)} = 1.4$ 
from an initial 2d square lattice with periodic boundary conditions and 
$d_{\mathrm{s}}^{(2)} = 1.1$ from a connected random graph $G(N,M)$.  One notices that 
after $10^6$ steps, the logarithmically integrated DOS follows the respective target, 
$\tilde{I}^{\textrm{target}}(\ln\tilde{\lambda}) = d_{\mathrm{s}}/2 \, \ln\tilde{\lambda}$, 
very closely up to immanent discretization.

\begin{figure}[!htb]
  \centering
  \includegraphics[width=\linewidth]{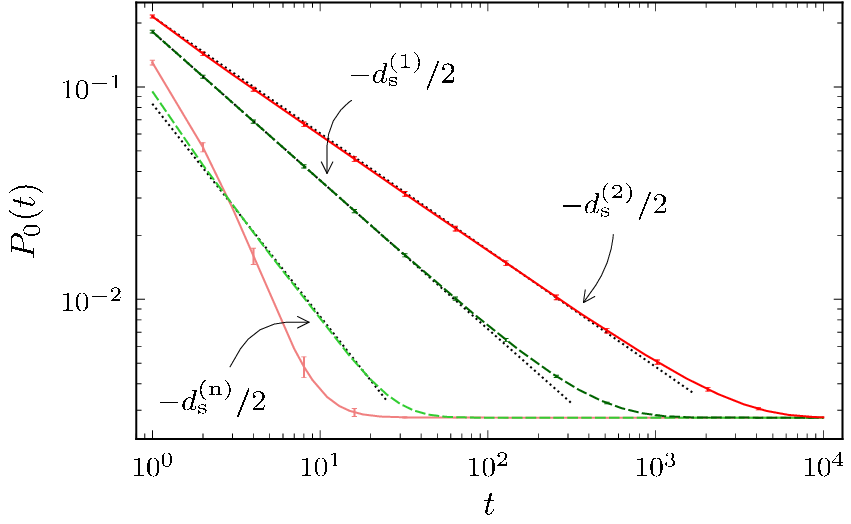}
  \caption{(Color online) Average probability~$P_0(t)$ 
    for a random walker to return to its origin at time~$t$
    on the initial networks (light colors) and 
    evolved networks after $10^6$ steps 
    with power-law DOS~$\tilde{I}^{\textrm{target}}(\ln\tilde{\lambda}) = d_{\mathrm{s}}/2
    \, \ln\tilde{\lambda}$ as targets (dark colors) for 
    evolutions with $d_{\mathrm{s}}^{(1)} = 1.4$ starting 
    from 2d square lattices 
    (green dashed lines)
    and $d_{\mathrm{s}}^{(2)} = 1.1$ starting from 
    random graphs (red solid lines).  The lines display the
    mean of $100$ realizations and the error bars mark one standard deviation of the
    distributions.  The black dotted lines are guides to the eye and have slopes of
    $-d_{\mathrm{s}}^{\mathrm{(n)}}/2 = -1$,
    $-d_{\mathrm{s}}^{(1)}/2 = -0.7$, and $-d_{\mathrm{s}}^{(2)}/2 = -0.55$, as annotated,
    indicating the expected scaling behavior.
  }
  \label{fig:rwPorigin}
\end{figure}
The average probability~$P_0(t)$ for a random walker to return to its origin at 
time~$t$ is obtained for the initial as well as for the evolved networks and is 
shown in fig.~\ref{fig:rwPorigin}.  For the evolved networks, $P_0(t)$ indeed 
follows a power-law $P_0(t) \propto t^{-d_{\mathrm{s}}/2}$ with $d_{\mathrm{s}}^{(1)} = 1.4$ 
and $d_{\mathrm{s}}^{(2)} = 1.1$, respectively, as expected from the linear scaling of 
$\tilde{I}(\ln\tilde{\lambda})$ in fig.~\ref{fig:score_snapshot}.  Note the rather 
small dynamical heterogeneity indicated by the error bars (displaying one standard 
deviation).

\begin{figure}
  \centering
  \includegraphics[width=\linewidth]{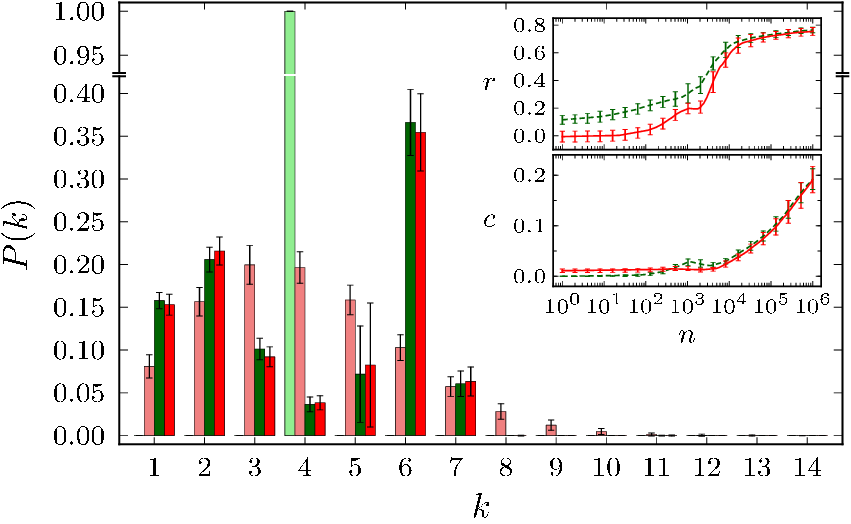}
  \caption{(Color online)%
    Degree distributions~$P(k)$ of---from left to right---the two initial and the 
    corresponding evolved networks after $10^6$ evolution steps, i.e., 2d square 
    lattices (1st bar, light green), random graphs with mean 
    $\langle k \rangle = 4$ (2nd bar, light red), and the resulting networks after 
    evolution towards a power-law DOS~$\tilde{I}^{\textrm{target}}(\ln\tilde{\lambda}) =
    d_{\mathrm{s}}/2 \, \ln\tilde{\lambda}$ with $d_{\mathrm{s}}^{(1)} = 1.4$ as target
    (3rd bar, dark green, and 4th bar, dark red).  The histogram bars display the 
    mean of $100$ networks, and the error bars mark one standard deviation.  
    The insets display the distribution of Newman coefficients~$r$ and clustering 
    coefficients~$c$ as a function of the step number~$n$ for evolutions starting 
    from 2d square lattices (green dashed lines) and random graphs (red solid 
    lines).  The lines display the averages over $100$ networks, and the error bars 
    mark one standard deviation of the distributions.
  }
  \label{fig:degree+newman}
\end{figure}
Given the convergence of the dynamical performance, the question arises as to how similar 
the underlying networks are topologically.  In fig.~\ref{fig:degree+newman}, the 
degree distributions~$P(k)$, being the probability that a randomly chosen vertex has 
degree~$k$, of the two initial and the corresponding evolved networks are shown.  
Despite the dissimilarities in the degree distributions of the two classes of initial 
networks (Poisson vs.\ delta distribution), the degree distributions of the evolved 
networks are quite alike on the level of averages.  Nevertheless, there is significant 
variation from network to network as can be seen, for example, by the large standard 
deviation for $k=5$ indicated by the error bars.  While very few vertices with the 
average degree $k=4$ are present, the fraction of vertices with higher ($k=6$) and 
lower ($k=1,2$) degrees increases in the course of the evolution suggesting that the 
networks become more heterogeneous.

To gain further insight into the resulting network topologies, we display in the
insets of fig.~\ref{fig:degree+newman} the average degree assortativity or Newman 
coefficient~$r$ and the average clustering coefficient~$c$ as function of the evolution 
step number~$n$.  The Newman coefficient~$r$~\cite{newman_assortative_2002} is the 
Pearson correlation coefficient of degrees of two vertices connected by an edge and 
hence measures the amount of degree-degree correlation (degree assortativity) in a 
network.  A network without degree-degree correlations has $r = 0$.  In our example, 
the networks become assortative in the course of the evolution ($r > 0$), meaning that 
connections preferentially exist between vertices of similar degree, another indicator 
for heterogeneity in the network structure.  The clustering coefficient~$c$, defined as 
the density of triangles~\cite{barrat_properties_2000}, measures the transitivity of a 
network, i.e., the tendency that the neighbors of a given vertex are connected to one 
another.  For a connected graph with fixed numbers of vertices and edges, an increasing 
clustering coefficient $c$ indicates, again, a trend towards more heterogeneous 
structures.  We observe that although the network evolution converged in less than 
$10^6$ evolution steps with respect to the dynamical behavior and the Newman coefficient, 
the clustering coefficient is still increasing and broadly distributed.  As the changes 
in the different observables do not coincide with the changes in the dynamical behavior, 
we conclude that even with a given eigenvalue spectrum there is some structural freedom 
left for variation in 3-point properties such as transitivity and, despite the dynamical 
behavior being essentially identical after $10^6$ evolution steps, the underlying network 
topologies can still be rather diverse.  We do not observe any systematic differences 
between networks evolved from regular lattices or random graphs suggesting that already 
this simplest version of our evolutionary algorithm suffices to explore the relevant 
region of the network configuration space in a reasonable number of evolution steps.

\section{Conclusion}  Most networks, in particular as occurring in biology, are not 
static entities but instead constantly evolve over time.  Even though a given change 
usually affects the network structure in the first place, its fate is decided by the 
resulting alteration of the dynamics on the network providing its functionality, so 
that evolution favors topologies with a better dynamical performance.  We study the 
relationship between network structure and dynamics by means of network evolution 
using a performance measure based on the whole eigenvalue spectrum of the time evolution 
operator of linear dynamics.  By this, we avoid the explicit calculation of the 
dynamical time evolution, which can be computationally very costly.  To exemplify our 
approach, we use the graph Laplacian describing a diffusion process, and evolve networks 
such that a given sub-diffusive behavior emerges.  Interestingly, the resulting networks 
show an overall trend towards heterogeneous structures and, despite displaying 
essentially the same dynamical behavior, are structurally quite diverse.

\section{Acknowledgment}  We gratefully acknowledge helpful discussions with
Sebastian Weber in early stages of this work and funding by the
Stu\-dien\-stif\-tung des deut\-schen Vol\-kes and the
Bonn-Cologne Graduate School of Physics and Astronomy (SK) and by the
Deut\-sche For\-schungs\-ge\-mein\-schaft via the Heisen\-berg program
(PO~1025/6) (MP).


\end{document}